\documentclass[preprint]{aastex}
\usepackage{natbib}
\usepackage{booktabs}
\usepackage{threeparttable}

\newcommand{\fexxi}{Fe \scriptsize{XXI} \normalsize}
\newcommand{\ci}{C \scriptsize{I} \normalsize}
\newcommand{\oi}{O \scriptsize{I} \normalsize}
\newcommand{\fexix}{Fe \scriptsize{XIX} \normalsize}
\newcommand{\fexii}{Fe \scriptsize{XII} \normalsize}
\newcommand{\fexiii}{Fe \scriptsize{XIII} \normalsize}
\newcommand{\feii}{Fe \scriptsize{II} \normalsize}
\newcommand{\siii}{Si \scriptsize{II} \normalsize}
\newcommand{\sxv}{S \scriptsize{XV} \normalsize}
\newcommand{\caxix}{Ca \scriptsize{XIX} \normalsize}

\begin{document}

\title{Doppler shift oscillations from a hot line observed by IRIS}

\author{D.~Li\altaffilmark{1,2,3}, Z.~J.~Ning\altaffilmark{1}, Y.~Huang\altaffilmark{1}, N.-H.~Chen\altaffilmark{4}, Q.~M.~Zhang\altaffilmark{1}, Y.~N.~Su\altaffilmark{1}, and W.~Su\altaffilmark{1,2}}
\affil{$^1$Key Laboratory of Dark Matter and Space Astronomy, Purple Mountain Observatory, CAS, Nanjing 210008, China \\
    $^2$Key Laboratory of Modern Astronomy and Astrophysics (Nanjing University), Ministry of Education, Nanjing 210023, China \\
    $^3$CAS Key Laboratory of Solar Activity, National Astronomical Observatories, Beijing 100012, China \\
    $^4$Korea Astronomy and Space Science Institute, Daejeon 34055, Korea}
\altaffiltext{}{Correspondence should be sent to: lidong@pmo.ac.cn,
ningzongjun@pmo.ac.cn}

\begin{abstract}
We present a detailed investigation of the Doppler shift
oscillations in a hot loop during an M7.1 flare on 2014 October 27
observed by the {\it Interface Region Imaging Spectrograph}. The
periodic oscillations are observed in the Doppler shift of
\fexxi~1354.09~{\AA} (log$T\sim$7.05), and the dominant period is
about 3.1~minutes. However, such 3.1-min oscillations are not found
in the line-integrated intensity of \fexxi~1354.09~{\AA}, AIA EUV
fluxes, or microwave emissions. {\it SDO}/AIA and {\it Hinode}/XRT
imaging observations indicate that the Doppler shift oscillations
locate at the hot loop-top region ($\geq$11~MK). Moreover, the
differential emission measure (DEM) results show that the
temperature is increasing rapidly when the Doppler shift oscillates,
but the number density does not exhibit the corresponding increases
nor oscillations, implying that the flare loop is likely to
oscillate in an incompressible mode. All these facts suggest that
the Doppler shift oscillations at the shorter period are most likely
the standing kink oscillations in a flare loop. Meanwhile, a longer
period of about 10~minutes is identified in the time series of
Doppler shift and line-integrated intensity, {\it GOES} SXR fluxes
and AIA EUV light curves, indicating the periodic energy release in
this flare, which may be caused by a slow mode wave.
\end{abstract}


\section{Introduction}
Oscillations have been commonly detected in the solar atmosphere,
and they are often characterized by the regular peaks or the
periodic motions. The oscillations can be identified as the
brightness fluctuations from the total fluxes in a broad wavelength,
such as microwave emissions \citep{Melnikov05,Tan12},
extreme-ultraviolet and ultraviolet (EUV/UV) passbands
\citep{Dem02,Ning17}, soft and hard X-ray (SXR/HXR) channels
\citep{Lipa78,Li08,Ning14}. On the other hand, the oscillations can
also be identified as the spatial displacement movements in the
imaging observations \citep{Aschwanden99,Shen12,Shen13}, as well as
in the spectroscopic observations such as the properties of the line
profiles $-$ Doppler shift, line intensity and width
\citep{Tian11,Lit15,Li15a,Zhang16a}. The current discovered
oscillatory periods can range from tens of milliseconds
(10$^{-2}$~s) to thousands of seconds (10$^{3}$~s) in the
multi-wavelength observations from radio through EUV/UV to X-ray
emissions \citep[e.g.,][]{Schrijver02,Inglis09,Dem12}. And the
oscillations within multi-periods have also been detected in the
same event \citep{Zimovets10,Yang16,Li17a}.

Over decades, the oscillations in coronal or flare loops are
particularly investigated
\citep[e.g.,][]{Nakariakov99,Kleim02,Goddard16,Yuan16a}. Using the
imaging observations from the {\it Transition Region and Coronal
Explorer} ({\it TRACE}), the loop oscillations are first detected as
the spatial displacement oscillations in solar corona
\cite[see.,][]{Aschwanden99,Nakariakov99}, and details are
subsequently reported by \cite{Aschwanden02} and \cite{Schrijver02}.
In the {\it Solar Dynamics Observatory} ({\it SDO}) era, these
oscillations are further studied in flare or coronal loops
\citep[e.g.,][]{Wang12,Nistico13,Yuan16b}. Currently, two regimes
are found: one is the rapidly damping oscillations within only a few
cycles and the other is decayless oscillations. The amplitudes of
the rapidly damping oscillations are usually large and decay with
time \citep{Aschwanden99,Nakariakov99,Zimovets15}. While those
decayless oscillations with small amplitudes can last for tens of
cycles \citep{Nistico13,Anfinogentov15}. The loop oscillations can
also be observed in the warm and hot emission lines which formed in
solar corona or flares, such as \fexix 1118~{\AA} and \fexxi
1354~{\AA} from {\it SOHO}/SUMER \citep{Wang02,Wang03}, \caxix
3.1781~{\AA} and \sxv 5.0379~{\AA} from {\it YOHKOH}/BCS
\citep{Mariska05,Mariska06}, \fexii 195.12~{\AA} and \fexiii
202.04~{\AA} from {\it Hinode}/EIS \citep{Tian11}, \fexxi
1354.09~{\AA} from {\it IRIS} \citep{Tian16}.

Those oscillations detected in coronal or flare loops are often
explained by MHD waves \citep{Nakariakov09,Dem12}, such as fast kink
waves \citep{Su12a,Nistico13,Yuan16c}, slow magnetoacoustic waves
\citep{Ofman02,Wang03}, and fast sausage waves \citep{Su12b,Tian16}.
A remarkable application, namely MHD coronal seismology, derived
from those observational parameters is to diagnose the properties of
coronal plasma along the loops as well as the magnetic field
strength \citep[e.g.,][]{Nakariakov05,Dem12,Yuan16a}. This can also
greatly improve our current understanding of coronal heating and
magnetic reconnection theory \citep{Nakariakov99,Goossens13}.

Most recent studies of loop oscillations only focus on the imaging
observations \citep[e.g.,][]{Anfinogentov15,Zimovets15,Goddard16}
and here we utilize the spectroscopic data to measure the Doppler
shift oscillations on the \fexxi1354.09~{\AA} line which is formed
at the temperature of $\sim$11~MK. Our results provide strong
evidences for fast kink oscillations in a flare loop. The
observations are taken from the joint observations, including the
{\it Interface Region Imaging Spectrograph} \citep[{\it
IRIS},][]{Dep14}, the Atmospheric Imaging Assembly
\citep[AIA,][]{Lemen12} and Helioseismic and Magnetic Imager
\citep[HMI,][]{Schou12} on board {\it SDO}, the X-ray Telescope
\citep[XRT,][]{Golub07} aboard {\it Hinode}, the Nobeyama
Radioheliograph \citep[NoRH,][]{Hanaoka94}, and the {\it
Geostationary Operational Environment Satellites} \citep[{\it
GOES},][]{Aschwanden94}.

\section{Observations and Data Analysis}
\begin{figure}
\epsscale{0.8} \plotone{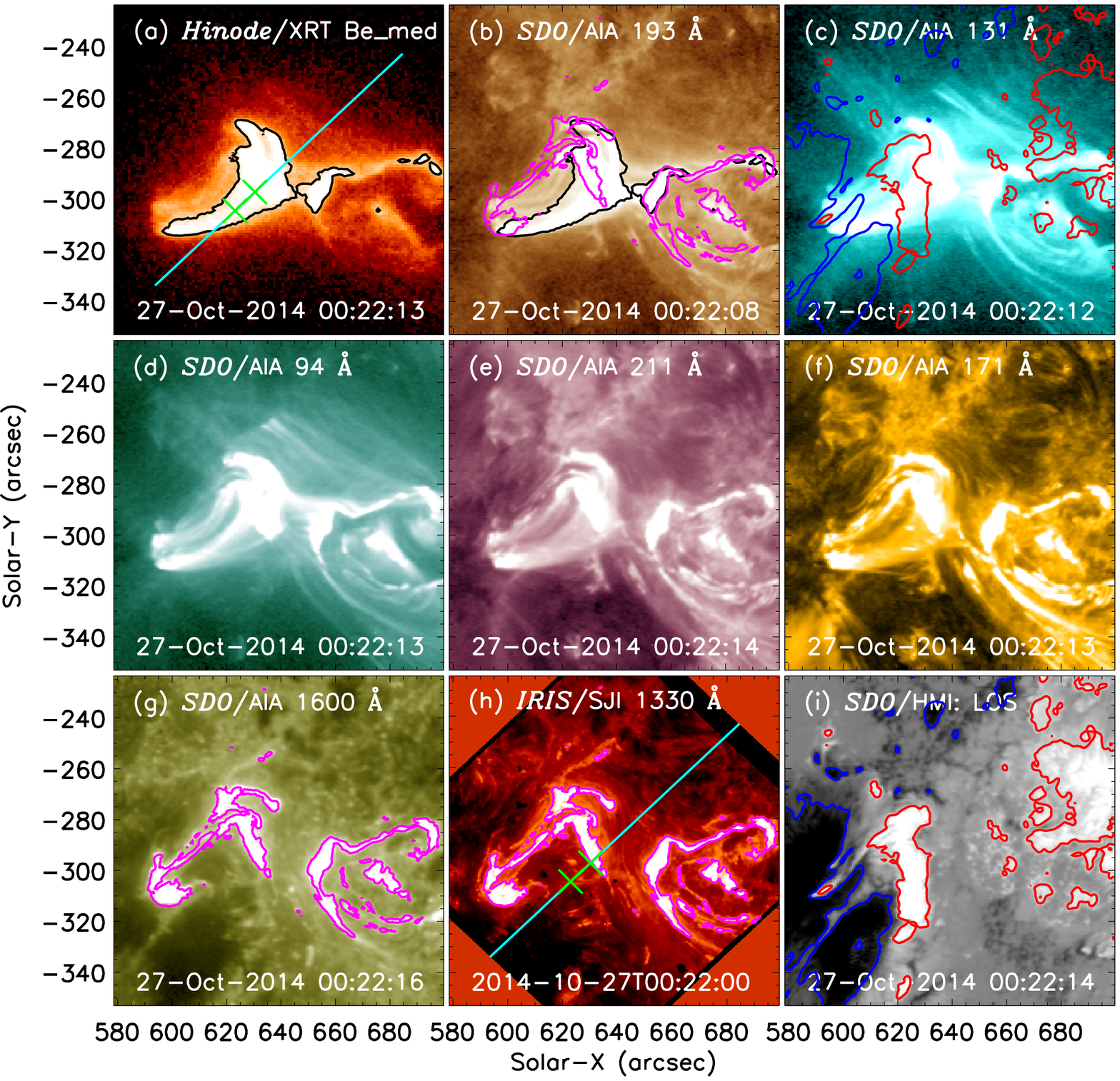} \caption{Simultaneous snapshots
observed by {\it Hinode}/XRT, {\it SDO}/AIA, {\it SDO}/HMI and {\it
IRIS}/SJI on 2014 October 27. The turquoise line marks the {\it
IRIS} slit, and two short green lines are the bounds of the selected
loop-top region. The black contours outline the SXR flare loop based
on the XRT Be\_med image with an intensity level of 700~DN, and the
purple contours show the flare ribbons observed in SJI 1330 {\AA}
image with an intensity level of 400 DN. The red and blue contours
represent the positive and negative fields at the scale of
$\pm$600~G from HMI LOS magnetogram. A logarithmic scale is used in
the XRT, AIA and SJI images.} \label{image}
\end{figure}

The M7.1 flare under our study occurs in AR 12192 on 2014 October
27, and its derived position is S12W42. It starts at about 00:06~UT,
peaks nearly at 00:34~UT, and stops at around 00:44~UT.
Figure~\ref{image} shows the simultaneous snapshots with the same
field-of-view (FOV) of 120$\arcsec$$\times$128$\arcsec$ at SXR,
EUV/UV and magnetogram observed by {\it Hinode}/XRT
($\sim$1\arcsec/pixel), {\it SDO}/AIA ($\sim$0.6\arcsec/pixel), {\it
IRIS}/SJI ($\sim$0.33\arcsec/pixel), and {\it SDO}/HMI
($\sim$0.6\arcsec/pixel), respectively. The XRT, AIA, and HMI images
have been pre-processed with the routines of `xrt\_prep.pro'
\citep{Golub07}, `aia\_prep.pro' \citep{Lemen12}, and
`hmi\_prep.pro' \citep{Schou12} in the solar software (SSW) package,
respectively. First, to coalign the XRT and AIA imaging
observations, we use Be\_med filter and 193 {\AA} images in which a
hot flare loop (indicated by the black contours) is clearly visible
and can be identified easily. Second, the AIA 1600~{\AA} image is
applied to co-align with the SJI 1330~{\AA} image
\citep{Li14,Cheng15}, because they both contain the continuum
emission from the temperature minimum (marked by the purple
contours). A broad and bright loop is visible in the XRT Be\_med
filter, AIA 193~{\AA} and 131~{\AA} images, and the legs of loop are
rooted individually in two polarities (see the blue and red
contours). The loop structure is weaker in other AIA channels, such
as the tip (or cusp-like tip) of the loop that is hardly seen in AIA
94 {\AA} and invisible in AIA 211~{\AA} and 171 {\AA}. The flare
ribbons are best shown in AIA 1600~{\AA} and SJI 1330~{\AA}. These
observations suggest that the loop structure, especially the loop
top, is very hot, which can be greater than 11~MK.

In order to investigate the loop oscillations carefully, we select
an area on the loop-top region that is crossed by the {\it IRIS}
slit. It is bounded by two short green lines in Figure~\ref{image}.
In this observation, the flare is observed in a `sit-and-stare' mode
with a step cadence of $\sim$16.2~s, which provides an opportunity
to study the evolution of flare spectra. First, two routines in the
SSW package, namely `iris\_orbitval\_corr\_l2.pro'
\citep{Tian14,Cheng15} and `iris\_prep\_despike.pro' \citep{Dep14}
are applied to the observed spectra. And here a relatively strong
neutral line (i.e., \oi 1355.60~{\AA}) are used to process the
absolute wavelength calibration \citep[see.,][]{Dep14,Tian15,Li17b}.
The multi-Gaussian functions superimposed on a linear background are
applied to fit the observed spectra. In the final step, the hot line
of \fexxi~1354.09~{\AA} can be extracted from the fitting result.
The detailed description of the fitting method can be seen in our
previous papers \citep{Li15b,Li16a}.

\begin{figure}
\epsscale{0.8} \plotone{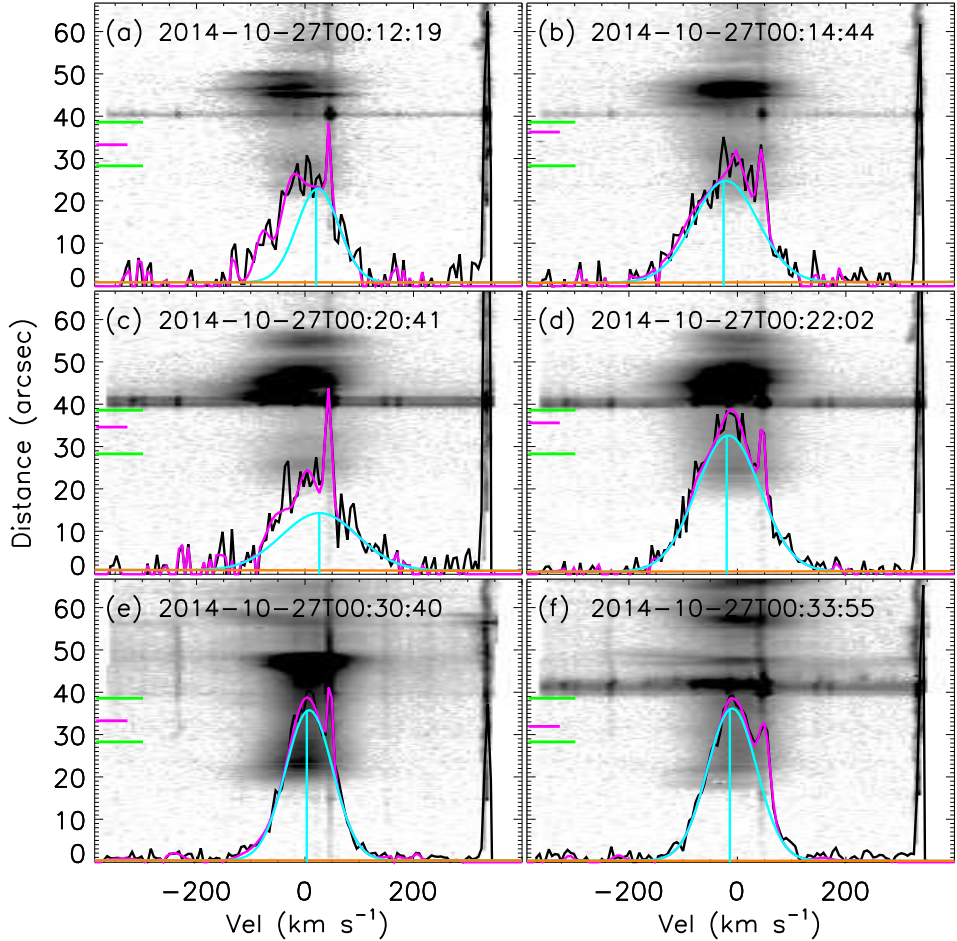} \caption{{\it IRIS} spectra and
their fitting results at `\oi' window. Two short green lines are the
bounds of the selected loop-top region. The black profile is the
observed {\it IRIS} spectrum at the slit position marked by the
short purple line. The purple profile gives the multi-Gaussian
fitting with a linear background (marked by the orange line), and
the turquoise profile is the extracted \fexxi~1354.09~{\AA} line.
The turquoise vertical line indicates the fitting line center of
\fexxi~1354.09~{\AA}.} \label{spe}
\end{figure}

Figure~\ref{spe} shows the flare spectra and the fitting results at
six selected time. It is well known that \fexxi 1354.09~{\AA} is a
broad and hot ($\sim$11~MK) line, whereas it is always blended with
other narrow and cool emission lines, such as \ci 1354.29~{\AA},
\feii 1353.02, 1354.01 and 1354.76~{\AA}, \siii 1352.63 and
1353.72~{\AA}, and some unidentified emission lines
\citep{Doschek75,Mason86,Innes03,Liy15,Young15}. Here, the flare
spectrum (black) is taken from the loop-top position indicated by a
short purple line. The purple overlaid profile represents the
multi-Gaussian fitting result, and the orange horizontal line is a
linear background. The overlaid turquoise profile is the
\fexxi~1354.09 {\AA} line extracted from the fitting result. The
other fitting parameter such as the line center is also overplotted
in each subfigure (vertical turquoise line) and used to calculate
the Doppler velocity \citep{Li15b}.

\section{Results}
\subsection{Doppler shift oscillations}
\begin{figure}
\epsscale{0.8} \plotone{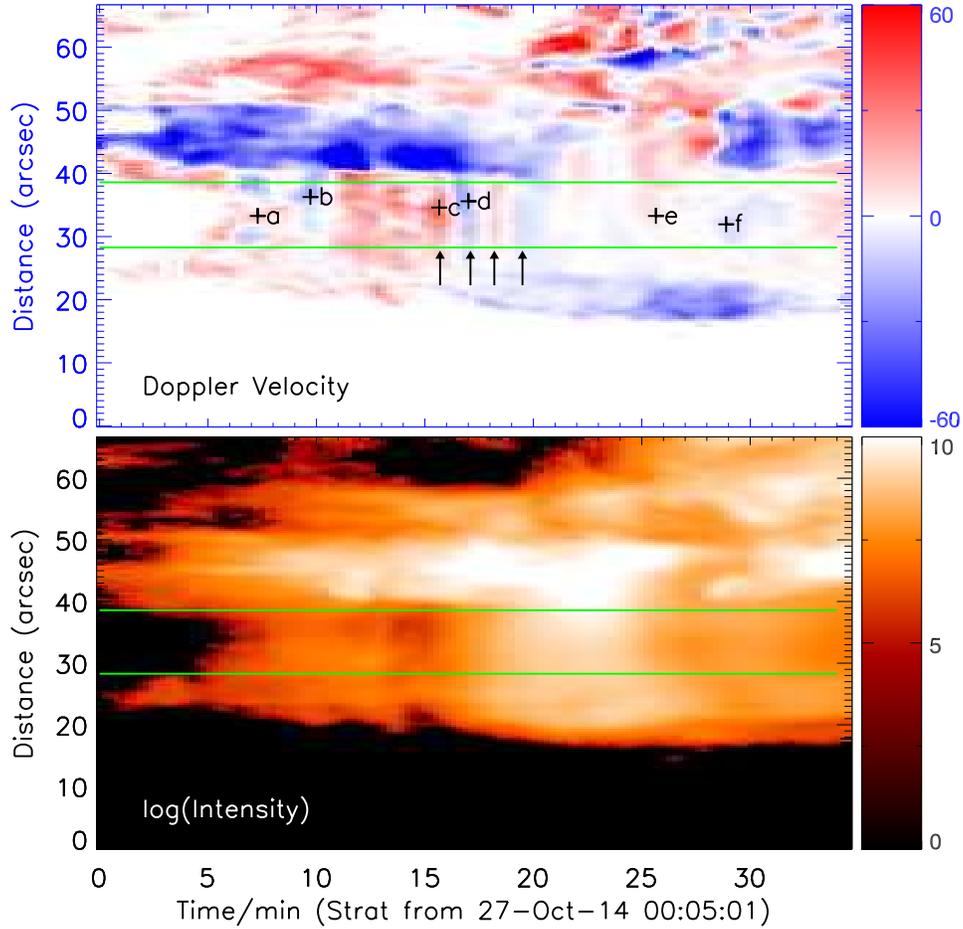} \caption{Upper: The space-time image
of Doppler shift at \fexxi 1354.09~{\AA}. Lower: The space-time
image of line-integrated intensity with a logarithmic brightness
scale. Two green lines are the bounds of the selected loop-top
region. The pluses (`+') mark the positions where the flare spectra
are shown in Figure~\ref{spe}.} \label{vel1}
\end{figure}

Figure~\ref{vel1} shows the space-time images of Doppler velocity
(upper) and line-integrated intensity (lower) at \fexxi
1354.09~{\AA}. The selected loop-top area is bounded by two green
lines, and the plus signs (`+') mark the time of the flare spectra
in Figure~\ref{spe}. A pronounced Doppler shift oscillation is
observed in the loop-top area from about 00:20~UT to 00:25~UT. It
begins from a red-shifted velocity, and then changes to a
blue-shifted velocity, such oscillatory behavior repeats at least
two times, as indicated by the solid arrows in the upper panel. In
other words, the oscillatory behaviors last for at least two cycles
if we define one cycle as a pair of red and blue shifts. After two
visible changes from red to blue shifts, the Doppler shifts become
very weak. However, no such obvious oscillations can be seen in the
space-time image of line-integrated intensity during the same time
intervals, as seen in the lower panel.

\begin{figure}
\epsscale{0.8} \plotone{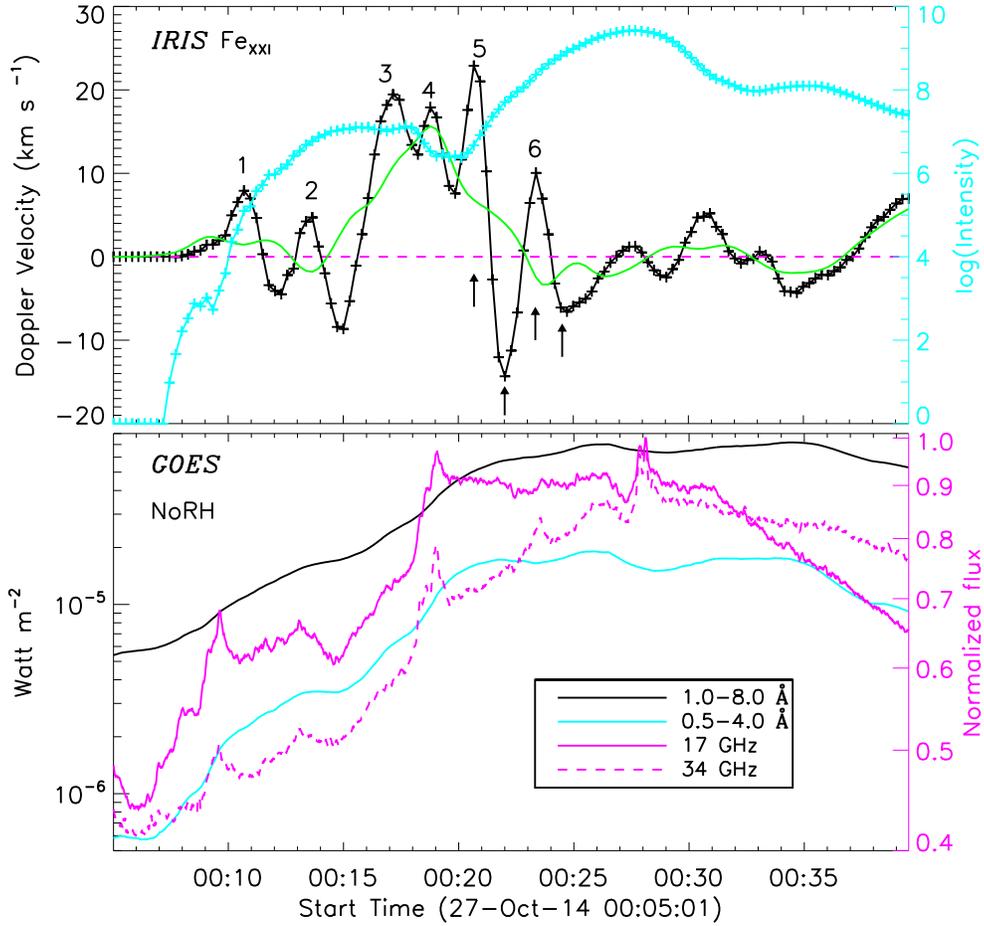} \caption{Upper: Temporal evolutions
of Doppler shift (black) and line-integrated intensity (turquoise)
at \fexxi~1354.09~{\AA}. The green profile represents the background
emission from Doppler shift. The oscillatory peaks are labeled with
the number ticks. The purple dashed line marks the zero velocity,
and the arrows mark the same time in Figure~\ref{vel1}. Lower: {\it
GOES} SXR light curves at 1.0$-$8.0~{\AA} (black) and
0.5$-$4.0~{\AA} (turquoise), and normalized fluxes at NoRH 17~GHz
(solid purple) and 34~GHz (dashed purple).} \label{flux}
\end{figure}

The upper panel in Figure~\ref{flux} shows the time series of
Doppler shift (black) and line-integrated intensity (turquoise),
they are averaged over the {\it IRIS} slit positions between two
green lines in Figure~\ref{vel1}. As a comparison, the SXR light
curves at 1.0$-$8.0 {\AA} (black) and 0.5$-$4.0 {\AA} (turquoise)
from {\it GOES} and normalized microwave fluxes at 17 GHz (solid
purple) and 34 GHz (dash purple) from the NoRH are also given in the
lower panel of Figure~\ref{flux}. The time series of Doppler shift
exhibit six pronounced and regular peaks between around
00:10~UT$-$00:25~UT, which are labeled by the number ticks. The
oscillations during the peaks `5' and `6' exhibit the changing from
red to blue wings, as marked by the solid arrows. The same
oscillations can be observed during the peaks `1' and `2', which
also display a change from red to blue wings. Hoverer, there are
only red-shifted oscillations during the peaks `3' and `4', and they
are always dominated by red wings. It is also noted that the onset
of these Doppler shift oscillations occurs in the impulsive phase of
this flare. On the other hand, none of the similar oscillatory
signature is revealed in the line-integrated intensity of
\fexxi~1354.09~{\AA}, normalized microwave fluxes and SXR light
curves. But the enhancement of line-integrated intensity is
simultaneous with the appearance of Doppler shift oscillations,
e.g., in peaks `1', and `5'. In peak `4', the line-integrated
intensity decreases when the Doppler shift tends to be red-shifted.
It is well known that the microwave emissions are produced by the
non-thermal electron beams trapped in a flare loop during the
impulsive phase of solar flare
\citep[see.,][]{Kundu94,White03,Reznikova11,Dolla12,Asai13,Kumar16}.
Although our observations does not show in-phase oscillations in the
normalized microwave fluxes at 17~GHz and 34~GHz, it is interesting
to note that the pulse peaks of microwave emissions proceed the
onset of Doppler shift oscillations by $\sim$1~minute, indicating
the non-thermal electrons produced by magnetic reconnection before
Doppler shift oscillations in this flare.

\begin{figure}
\epsscale{0.8} \plotone{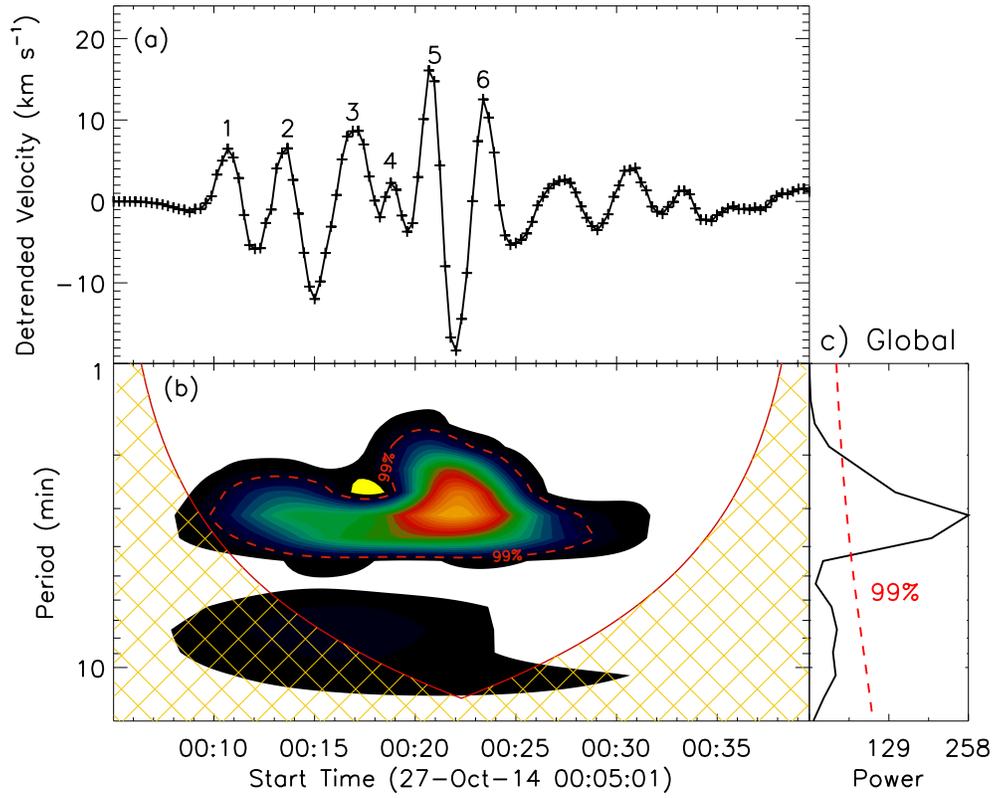} \caption{Panel~(a): Temporal
evolution of the detrended Doppler shift at \fexxi 1354.09~{\AA}.
Panel~(b) and (c): Wavelet power spectrum and global power of the
detrended Doppler shift. The dashed lines indicate a significance
level of 99\%.} \label{wave}
\end{figure}

To obtain the oscillatory period, a wavelet analysis method is used
\citep{Torrence98,Yuan11,Tian12,Li17c}. Firstly, a 5-min running
average in the time series of Doppler velocity is considered as a
background emission, as indicated by the green profile in the upper
panel of Figure~\ref{flux}. Secondly, the detrended time series of
Doppler velocity are calculated from the original time series by
removing the background emission (5-min running average).
Figure~\ref{wave}~(a) shows the detrended velocity, and they exhibit
six distinct peaks, which are well matched with those peaks in the
upper panel of Figure~\ref{flux}, as labeled by the number ticks.
Figure~\ref{wave}~(b) and (c) display wavelet power spectrum and
global wavelet, respectively. The oscillatory period can be measured
from the peak value of the global wavelet power, and the uncertainty
is thought to be the half width at half-maximum value
\citep{Tian16}. Finally, a dominant period with an error bar can be
estimated for the detrended Doppler velocity of
\fexxi~1354.09~{\AA}, which is about 3.1$\pm$0.6~minutes. The
wavelet power spectrum shows that these Doppler shift oscillations
are from $\sim$00:10~UT to $\sim$00:25~UT, which correspond to the
time interval during the six peaks in the time series of Doppler
velocity (see Figure~\ref{flux}).

\subsection{Flare loop geometry}
\begin{figure}
\epsscale{0.8} \plotone{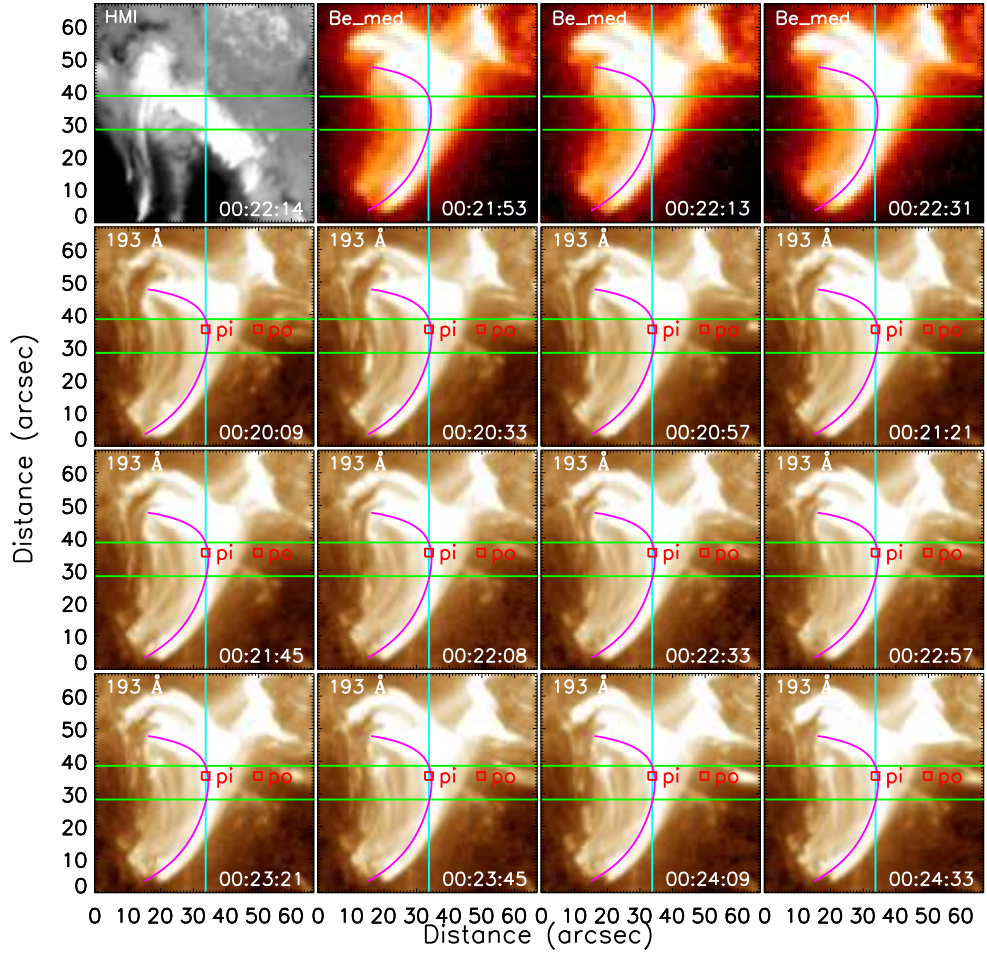} \caption{Snapshots along {\it IRIS}
slit direction observed by HMI, XRT, and AIA, respectively. Two
green lines are the bounds of the selected loop-top region, and the
turquoise line marks the {\it IRIS} spectral slit. The purple line
indicates the SXR loop. The red boxes outline the region
(2.4\arcsec$\times$2.4\arcsec) used to do the DEM analysis in
Figure~\ref{dem}. The scale levels of magnetogram are $\pm$1000~G,
and the XRT and AIA images are displayed on a logarithmic brightness
scale.} \label{loop}
\end{figure}

Thanks to the high spatial resolution of imaging observations, we
are able to study the morphological structure of the oscillatory
positions. Due to data gaps in XRT observations, only a few XRT
images are presented to show the flare loop structure at high
temperature during the time interval of the pronounced oscillations.
The SXR and EUV images are carefully selected to avoid data
saturation \citep[see also.,][]{Li15a,Ning17} and plotted with the
Y-axis along the direction of {\it IRIS} slit, e.g., 45 degree
rotation. Figure~\ref{loop} displays the loop structure and its
movement from peak~`5' to peak~`6' at XRT Be\_med filter and
AIA~193~{\AA}, it also shows the line-of-sight (LOS) magnetogram
with the same FOV. Figure~\ref{loop} shows that double footpoints
which connected by this flare loop root in the positive and negative
magnetic fields, respectively. We trace this flare loop in XRT
images and overlaid on each AIA~193~{\AA} image (purple line). The
region of interested is the loop-top area which is bounded by two
green lines here. The bright loop in AIA~193~{\AA} images is
cospatial with the overlaid SXR loop. It is noted that the bright
loop-top region (outlined with two green lines) is crossed by the
slit of {\it IRIS} (turquoise line).

Further, we also calculate some parameters of this flare loop. The
apparent loop length is estimated to be $\sim$54~Mm according to the
XRT images in Figure~\ref{loop}, as shown by the purple line. Then
we can obtain the deprojected loop length (L~$\approx$~81~Mm) when
considering the projection effect due to the derived position of
solar flare \citep{Aschwanden02,Kumar13,Li17d}. The loop width ($w$)
is also calculated from the XRT images. We first average the
intensities between two green lines and obtain the flux along the
X-direction. Then we fit the flux with a Gaussian function. Finally,
the full width at half maximum (FWHM) of Gaussian function is
considered as the loop width, which is about 9.8~Mm. Thus, the
geometric ratio between loop width and length ($w$/L) is around
0.12. This geometric ratio is quite reasonable, given $w=2 \times r$
(r is the radius of flare loop), therefore, the aspect ratio between
loop radius and length ($r$/L) is about 0.06, which is consistent
with previous findings in the coronal loops \citep{Aschwanden02}.

\subsection{Temperature and density}
\begin{figure}
\epsscale{0.8} \plotone{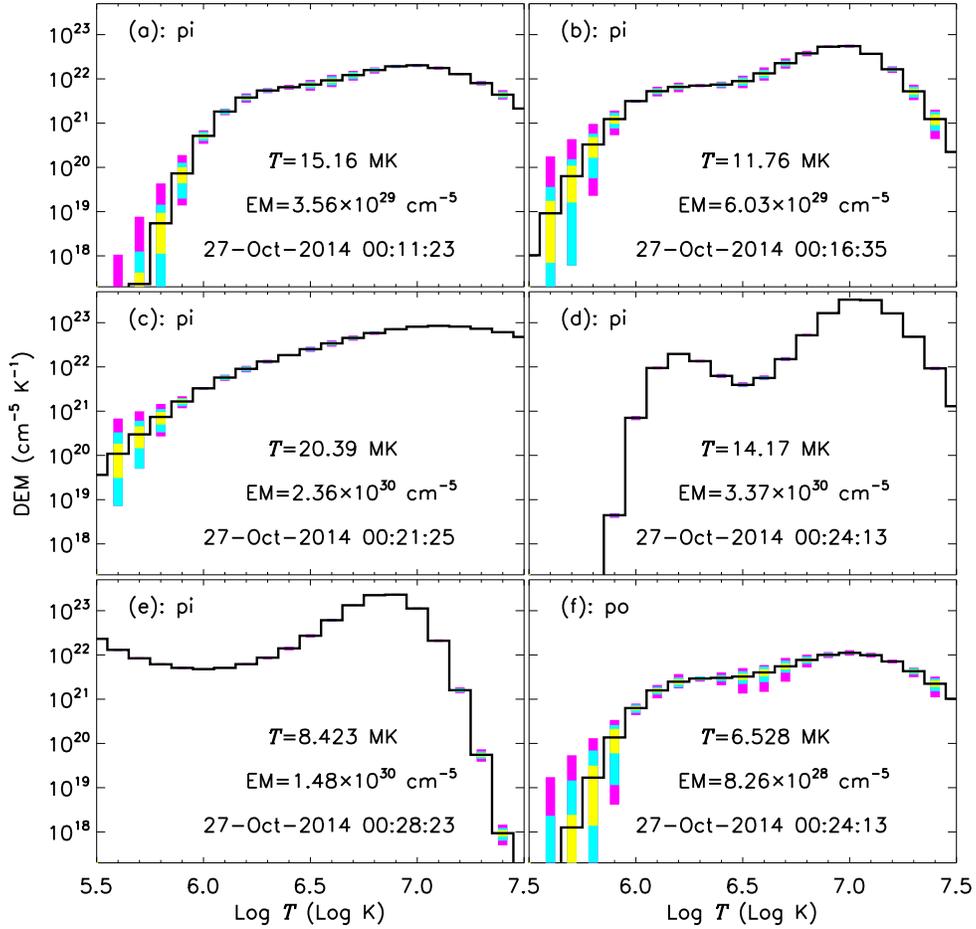} \caption{DEM profiles in the flare
loop top (a-e) and background (f) regions, which are outlined by the
small red boxes in Figure~\ref{loop}. The black profile shows the
best-fitted DEM curve from AIA EUV observations. The yellow
rectangles represent the regions that contains 50\% of the MC
solutions. The turquoise rectangles, above and below the yellow
rectangles, and the yellow rectangles compose the regions that cover
80\% of the MC solutions. All the colored rectangles form the
regions which contain 95\% of the MC solutions. The mean
temperature, EM and observed time are given in each panel.}
\label{dem}
\end{figure}

To investigate the temperature structure of this flare loop, we
perform the differential emission measure (DEM) analysis using AIA
data at six EUV wavelengths \citep[see.,][]{Cheng12,Sun14,Shen15}.
Figure~\ref{dem}~(a)$-$(e) shows the DEM profiles in the loop-top
region (marked by the red box of pi in Figure~\ref{loop}) with a FOV
of 2.4\arcsec$\times$2.4\arcsec\ at the selected time. While
Figure~\ref{dem}~(f) gives the DEM profiles outside flare loop, as
indicated by the red box of po. The average temperature ($T$) and
emission measure (EM) at each time are estimated within a confident
temperature (log$T$) range of 5.5$-$7.5 \citep{Ning16}. The
best-fitted DEM solution of the observation is indicated by the
black profile. Then the Monte Carlo (MC) realizations of the
observations are applied to estimate the DEM uncertainty, as
indicated by the top and bottom color rectangles
\citep[see.,][]{Cheng12,Li16b,Lil16c,Lil16d}. It gives much lower
value of the uncertainty on our DEM analysis, especially at the high
temperature regions, i.e., panels (c) and (d).

\begin{figure}
\epsscale{0.8} \plotone{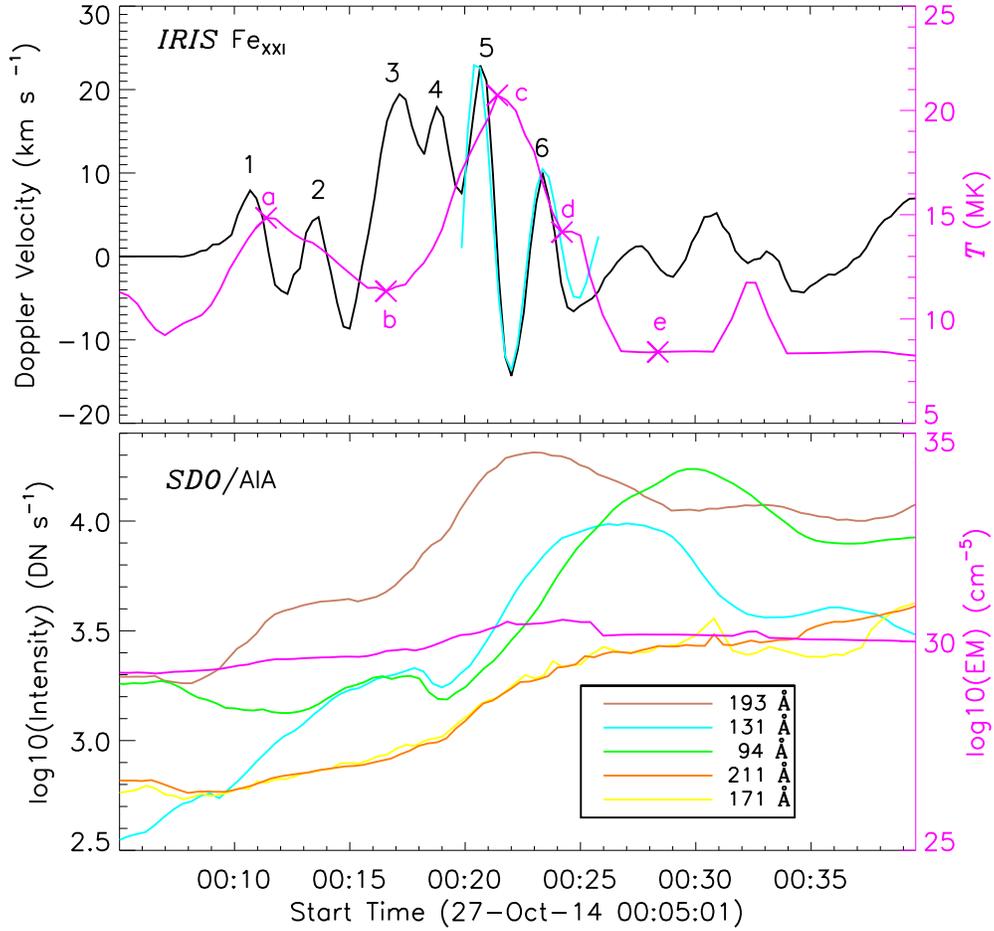} \caption{Upper: Time series of
Doppler shift at \fexxi 1354.09~{\AA} and temperature (purple) in
the flare loop-top region. The turquoise line gives the best fit for
the peaks `5' and `6'. The crosses (`$\times$') indicate the time to
display the DEM analysis results in Figure~\ref{dem}. Lower: {\it
SDO}/AIA light curves at 193~{\AA} (brown), 131~{\AA} (turquoise),
94~{\AA} (green), 211~{\AA} (orange), and 171~{\AA} (yellow). The
purple profile is the temporal variations of EM in the flare
loop-top region.} \label{vte}
\end{figure}

The upper panel of Figure~\ref{vte} is the time series of the
Doppler shift and temperature at the flare loop-top region, and the
lower panel shows the EM variation and EUV fluxes at AIA 193~{\AA}
($\sim$20~MK), 131~{\AA} ($\sim$11~MK), 94~{\AA} ($\sim$6.3~MK),
211~{\AA} ($\sim$2.0~MK), and 171~{\AA} ($\sim$0.63~MK)
\citep{Lemen12}. The temperature rises rapidly at the onset of
Doppler shift oscillations (e.g., peak `1' $\&$ `5'), which is
almost in phase with the enhancement of the EUV fluxes at AIA
131~{\AA} and 193~{\AA} passbands ($\geq$11~MK). The temperature
increases here might be due to hot emission from the loop top that
are clearly seen in the XRT Be\_med and AIA~193~{\AA} images
(Figures~\ref{loop}) and also crossed by the slit of {\it IRIS}.
However, the EUV fluxes in other channels lags by several minutes
and such in-phase correspondence can not be found, implying that the
Doppler shift oscillations only take place in the hot loop, i.e., at
least 11~MK. The EM varies very slowly throughout the entire
impulsive phase. It keeps nearly constant during the occurrence of
Doppler shift oscillations, suggesting the incompressible flare
loop. The average EM in the flare loop-top area between
$\sim$00:20~UT and $\sim$00:25~UT is estimated to be
$\sim$3$\times$10$^{30}$~cm$^{-5}$. We further estimate the average
number density from equation~\ref{dens} \citep[see
also.,][]{Zhang14,Su15,Ning16,Zhang16b}, which is about
5.5$\times$10$^{10}$~cm$^{-3}$. This value agrees well with previous
findings in flare loops \citep{Aschwanden97}, for example,
$n_e\approx$(2$-$25)$\times$10$^{10}$~cm$^{-3}$ in SXR loops.

\begin{equation}
    n_e=\sqrt{\frac{EM}{w}}
    \label{dens}
\end{equation}

\section{Discussions}
\subsection{Error analysis}
We notice that the Doppler velocities shown in Figure~\ref{flux} are
not pronounced. For example, they are no more than
$\pm$25~km~s$^{-1}$ during the Doppler shift oscillations. The
Doppler velocity in this study is calculated from the subtraction
between the fitting line center and the rest wavelength, and there
could be two quantities producing the errors. One is the fitting
line center derived from the multi-Gaussian fitting method. The
multi-Gaussian fitting procedure \citep[detail see.,][]{Li15b,Li16a}
used to obtain the \fexxi~1354.09~{\AA} here has included as much as
possible the known/unknown blending emission lines in this dataset.
The final fit as well as the extracted \fexxi~1354.09~{\AA} line
profile (see Figure~\ref{spe}) appear to be successful at the
selected positions. Thus the derived fitting line center is
convincing. The other one is the rest wavelength. The formation
temperature of \fexxi 1354.09~{\AA} is $\sim$11~MK, which is too hot
to be observed in the quiet-Sun spectrum. In this paper, we set the
rest wavelength as 1354.09~{\AA}, which is the average value from
the recent {\it IRIS} spectral observations, i.e.,
1354.08~{\AA}$-$1354.1~{\AA}
\citep[e.g.,][]{Polito15,Tian15,Sadykov15,Brosius16}. And the
uncertainty is around $\pm$0.01{\AA}, which corresponds to the
Doppler shift of about $\pm$2.2~km~s$^{-1}$. This uncertainty on
rest wavelength of \fexxi~1354.09~{\AA} might change the Doppler
shift slightly but does not have a significant impact on our
results. On the other hand, \cite{Tian16} find that the Doppler
velocity of \fexxi~1354.09~{\AA} in a flare loop is much smaller
than that in the flare ribbons, which is similar as our results.

\subsection{MHD waves}
The oscillations in flare/coronal loops are usually related to MHD
processes \citep{Roberts00,Dem12,Tian16}, such as slow
magnetoacoustic waves, fast sausage waves and fast kink waves. In
this paper, the loop oscillations are observed in Doppler shift of
\fexxi~1354.09~{\AA}, but there are no apparent corresponding
intensity fluctuations in line-integrated intensity of
\fexxi~1354.09~{\AA} (Figure~\ref{flux}) or AIA EUV fluxes
(Figure~\ref{vte}). In DEM analysis, the EM is also relatively
stable during the Doppler shift oscillations. These facts suggest
that the flare loop oscillates in an incompressible mode, and
further rule out the possibility of slow magnetoacoustic waves
because they are compressible magnetoacoustic waves
\citep{Roberts00,Wang02,Wang03,Aschwanden05}. The loop oscillations
are unlikely to be interpreted as fast sausage waves either. The
sausage oscillations within short periods ($<$60~s) are often
decayless or weakly damping when the cutoff condition and those
within long periods ($>$60~s) are substantially compressible and
transverse \citep{Gruszecki12,Vasheghani14,Tian16}. Moreover, they
could produce density variation, which can easily modulate the
periodic oscillations in microwave emission
\citep{Rosenberg70,Aschwanden05,Antolin13,Reznikova15}. The loop
oscillations here are incompressible, damping, and no corresponding
intensity fluctuations are detected in EUV and microwave emissions.
All those theoretical predictions of fast sausage waves argue
against our observations.

The observed loop oscillations are most probably fast kink waves.
The oscillations produced by kink waves can produce the intensity
and Doppler shift based on the column depth from different viewing
angles. If the spectral slit is aligned along the loop, only Doppler
shift oscillations are expected from the LOS direction
\citep{Yuan16a}. In addition, the kink oscillations can also decay
rapidly \citep{Nakariakov99,Aschwanden05,Zimovets15,Yuan16c}. All
these theories are consistent with our observations. In this event,
a flare loop is detected to be corresponded to the Doppler shift
oscillations. The observed oscillatory period in this flare loop is
$\sim$3.1$\pm$0.6~minutes. That is to say, the dominant period of
kink oscillations is $\sim$3.1~minutes. On the other hand, according
to the period (P$_k$) of the standing kink oscillations, we can
diagnose the magnetic field strength ($B$) in this flare loop from
equation~\ref{pk}
\citep{Roberts84,Nakariakov01,Wang02,Aschwanden05,Tian12}.

\begin{eqnarray}
  B \approx \frac{v_A}{2.18\times10^{11}}n_{e}^{\frac{1}{2}}, \quad
  v_A = v_k(\frac{2}{1+n_0/n_e})^{-\frac{1}{2}}, \quad
  v_k = \frac{2L}{P_k}. \quad
 \label{pk}
\end{eqnarray}
Here, $v_A$ is the local Alfv\'{e}n speed, $v_k$ is the phase speed
of the standing kink wave in its fundamental mode, $n_e$ and $n_0$
are the electron densities inside and outside flare loop,
respectively. As mentioned in section~3, the deprojected length (L)
of this flare loop (Figure~\ref{loop}) is $\sim$81~Mm, thus the kink
speed can be estimated to be $\sim$870~km~s$^{-1}$. Meanwhile, the
number density ($n_e$) inside flare loop can be estimated from
equation~\ref{dens}. However, it is impossible to find a loop width
($w$) outside flare loop. Therefore, the effective LOS depth
($l\sim\sqrt{H\pi r}$) is used to estimate the number density
outside flare loop region (po), which is $\sim$4$\times$10$^{10}$~cm
\citep{Zucca14,Su16}. Then we can obtain the density ratio ($n_0/n_e
\sim$ 0.03) between outside (po) and inside (pi) flare loop at about
00:24:13~UT, as seen in Figure~\ref{dem}~(d) and (f). The Alfv\'{e}n
speed can be estimated to be about 630~km~s$^{-1}$, and the magnetic
filed strength in this flare loop can be estimated to be $\sim$68~G,
which agrees with pervious results, i.e., 60$-$120~G in flare loop
\citep[see][]{Qiu09}. In our observations, the density ratio
($n_0/n_e$) around flare loop is less than the typical density ratio
(i.e., 0.1$-$0.5) around coronal loop \citep{Aschwanden05},
suggesting that the number density in flare loop is higher than that
in coronal loop.

\subsection{Fourier analysis}
The upper panel in Figure~\ref{flux} exhibits that there are at
least two oscillatory periods in the time series of Doppler shift of
\fexxi~1354.09~{\AA}. One is the shorter period ($\sim$3.1~minutes)
which has been obtained in section~3.1. The other one seems to be a
longer period, and the time series of line-integrated intensity also
display the similar longer period. To extract the longer period, we
perform a Fourier analysis using the method described by
\cite{Yuan11}, which is based on the Lomb-Scargle periodogram
technique \citep{Scargle82,Horne86}.

\begin{figure}
\epsscale{0.8} \plotone{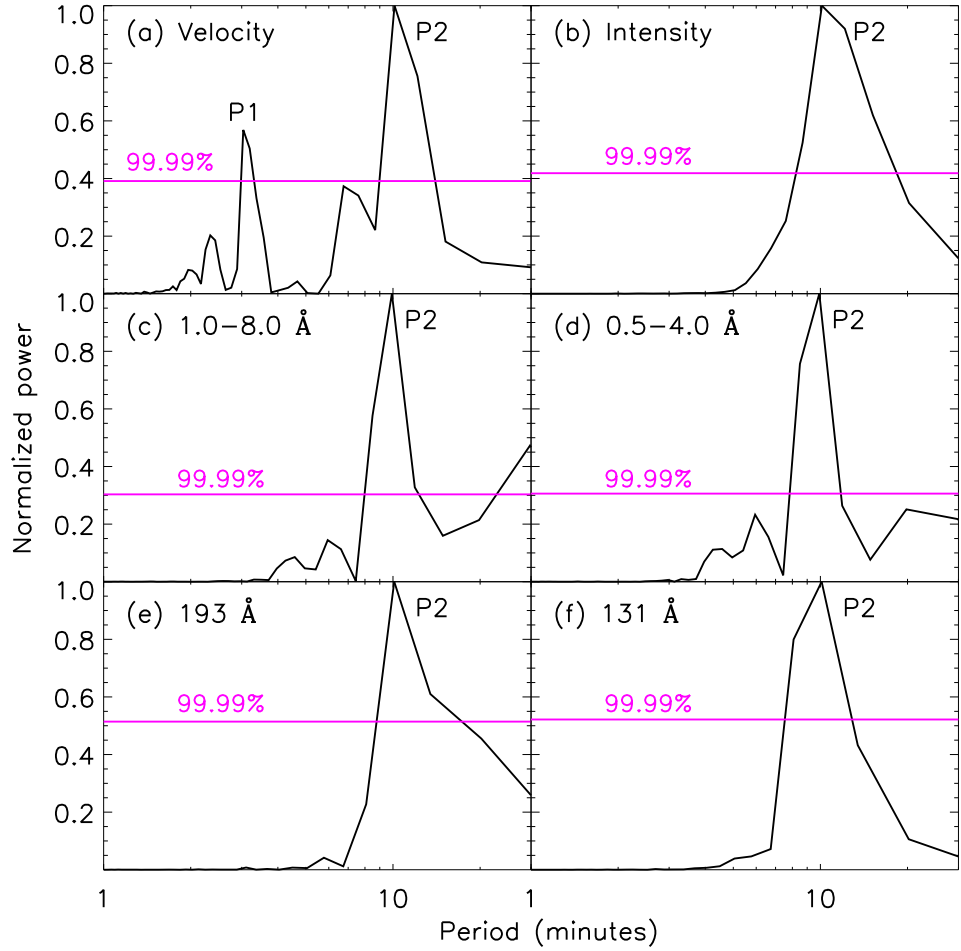} \caption{Normalized Fourier power
spectra of the detrended time series from {\it IRIS} \fexxi
1354.09~{\AA} (a, b), {\it GOES} SXR fluxes (c, d), and AIA EUV
light curves (e, f). A horizontal purple line in each panel
indicates the 99.99\% confidence level.} \label{fft}
\end{figure}

Figure~\ref{fft} shows the results of spectral analysis. The Fourier
analysis is applied to the detrended time series after removing the
background emission, which is the 12-min running average of the time
series \citep[see.,][]{Tian12,Li17c}. Here, a different averaging
time window only results in changing in relative power of the peaks
\citep{Yuan11}. Panel~(a) gives the normalized power spectrum of
detrended Doppler shift. From which, we can extract at least two
components of periods, P$_1$ and P$_2$. We also calculate the error
bars as the uncertainties of periods, which are the half width at
half-maximum value of the Fourier power \citep{Yuan11,Tian16}. The
shorter dominant period (P$_1$) within an error bar is around
3.1$\pm$0.6~minutes. The dominant period of P$_1$ agrees well with
the wavelet analysis result (Figure~\ref{wave}). While the longer
dominant period (P$_2$) within an error bar is about
10$\pm$4~minutes. This longer period can also be detected from the
dtrended time series of line-integrated intensity of
\fexxi~1354.09~{\AA} (b). To illustrate this longer period, we also
perform Fourier analysis of the detrended fluxes from {\it GOES}
1.0$-$8.0~{\AA} (c), 0.5$-$4.0~{\AA} (d) in SXR passbands, and AIA
193~{\AA} (f), 131~{\AA} (f) in EUV wavelengths. The normalized
power spectra show that all these detrended fluxes display a longer
dominant period of $\sim$10~minutes, which is same as that in
Doppler shift of \fexxi~1354.09~{\AA}. On the other hand, the
shorter period can not be detected in these detrended fluxes, which
further confirms that the 3.1-min oscillations can only be observed
in Doppler shift of \fexxi 1354.09~{\AA}.

As mentioned above, the Doppler shift oscillations within a period
of $\sim$3.1~minutes (P$_1$) can be interpreted as the standing kink
oscillations in the hot flare loop. On the other hand, the longer
period (P$_2$) is different from the shorter one (P$_1$), because it
can be observed simultaneously in SXR and EUV fluxes, as well as in
the time series of Doppler shift and line-integrated intensity of
\fexxi~1354.09~{\AA}. Those facts imply the process of periodic
energy release in this flare \citep{Tian15,Li17a}. The longer period
is most likely to be linked to the quasi-periodic pulsations (QPPs)
during the solar flare, which seems to be caused by a MHD wave in
slow mode \citep{Fang15,Van16}. Based on the formation temperature
($T \sim$11~MK) of \fexxi~1354.09~{\AA}, the local sound speed
($v_s$ $\approx$ 152$\sqrt{T/ \textbf{MK}}$) in flare loop is
estimated to be $\sim$500~km~s$^{-1}$
\citep[e.g.,][]{Nakariakov01,Kumar13,Kumar15}. While the deprojected
length (L) and longer period (P$_2$) of this flare loop have been
identified as $\sim$81~Mm and $\sim$10$\pm$4~minutes, then the phase
speed of this flare loop ($v_h$ = 2L/P$_2$) can be estimated as
about 200$-$450~km~s$^{-1}$. This phase speed is slower than the
local sound speed, implying it might be a slow MHD wave
\citep{Mandal16}. In other words, the longer period (P$_2$) might be
explained as the MHD wave in a slow mode.

To compare the Doppler shift oscillations, we also use the sine
function with a damping amplitude (equation~\ref{vfit}) to fit the
time series of Doppler shift during peaks `5' and `6'
\citep[e.g.,][]{Nakariakov99,Wang02,Anfinogentov15}, because the
Doppler shift oscillations are most pronounced during this time
interval. Figure~\ref{vte} (upper) shows the fitting result, the
overlaid turquoise profile gives the best fitting, and it appears to
match well with the Doppler velocities (black).

\begin{equation}
  v(t) = v_0 + v_m \sin(\frac{2\pi}{P_1} t + \psi) e^{-\frac{t}{\tau}}.
 \label{vfit}
\end{equation}
Here, $v_0$ is the initial Doppler velocity, $v_m$ is the initial
oscillatory amplitude, and P$_1$, $\psi$, and $\tau$ are the shorter
dominant period, initial phase, and decay time of the oscillations,
respectively.

\section{Conclusions}
Using observations from multi-instruments, i.e., {\it IRIS}, {\it
SDO}/AIA, {\it SDO}/HMI, {\it Hinode}/XRT and NoRH, we investigate
the Doppler shift oscillations in a hot loop during a {\it GOES}
M7.1 flare on 2014 October 27. The main results are summarized as
follows.

(I) The Doppler shift oscillations are detected in a hot flare line,
i.e., \fexxi 1354.09~{\AA} from {\it IRIS} spectroscopic
observations. The time series of Doppler shift start the
oscillations from red to blue wings (peaks `1' $\&$ `2'), and then
oscillate at red wings (peaks `3' $\&$ `4'), finally end the
oscillations from red to blue wings (peaks `5' $\&$ `6').

(II) Based on the wavelet analysis, a dominant period of
$\sim$3.1~minutes is obtained from the Doppler shift oscillations.
The Fourier analysis and sine function fitting provide an additional
evidence of this shorter period in the Doppler shift of
\fexxi~1354.09~{\AA}.

(III) The Doppler shift oscillations at the shorter period are from
the flare loop-top region. The temperature of the corresponding
flare loop is not less than 11~MK, which can be confirmed in the XRT
Be\_med filter, AIA~193 and 131~{\AA} images and DEM analysis in
section~3.3.

(IV) There are no apparent corresponding oscillations within the
period of $\sim$3.1~minutes in the line-integrated intensity of
\fexxi~1354.09~{\AA}, AIA EUV fluxes, or microwave emissions,
suggesting that the hot flare loop is incompressible. This is
consistent with the DEM analysis, which shows slightly change on EMs
during the Doppler shift oscillations.

(V) The 3.1-min period of Doppler shift oscillations at
\fexxi~1354.09~{\AA} might be explained by the MHD wave in a
standing kink mode. Thus, the magnetic filed strength in this flare
loop is estimated to be $\sim$68~G, and the density ratio between
outside and inside flare loop is $\sim$0.03.

(VI) The 10-min period oscillations can be observed in {\it GOES}
SXR fluxes, {\it SDO}/AIA EUV light curves and the time series of
Doppler shift and line-integrated intensity, implying the periodic
energy release in this flare, and it might be caused by a slow mode
MHD wave.

\acknowledgments The authors would like to thank the anonymous
referee for his/her valuable comments. We also acknowledge Prof.
H.~Tian, L.~P.~Li, D.~Yuan, P.~F.~Chen, X.~Cheng and Y.~Guo for
their inspiring discussions. We appreciate the teams of {\it IRIS},
{\it GOES}, NoRH, {\it Hinode}/XRT, {\it SDO}/AIA and {\it SDO}/HMI
for their open data use policy. This study is supported by NSFC
under grants 11603077, 11573072, 11773079, 11773061, 11473071,
11333009, XDA15052200, KLSA201708, 973 program (2014CB744200), and
Laboratory No. 2010DP173032. This work is also supported by the
Youth Fund of Jiangsu Nos. BK20161095, BK20171108, and BK20141043,
Dr. Q.~M.~Zhang is supported by the Surface Project of Jiangsu No.
BK 20161618 and the Youth Innovation Promotion Association CAS, and
Dr. Y.~N,~Su is also supported by one hundred talent program of
Chinese Academy of Sciences. The authors wish to thank the
International Space Science Institute in Beijing (ISSI-BJ) for
supporting and hosting the meeting of the International Team on
``Magnetohydrodynamic Seismology of the Solar Corona in the Era of
SDO/AIA", during which the discussions leading to this publication
were held.

\end{document}